# How Frequently are Articles in Predatory Open Access Journals Cited

Bo-Christer Björk,
Hanken School of Economics
Helsinki, Finland

Sari Kanto-Karvonen
Tampere University
Tampere, Finland

J. Tuomas Harviainen
Tampere University
Tampere, Finland

## Abstract

Predatory journals are Open Access journals of highly questionable scientific quality. Such journals pretend to use peer review for quality assurance, and spam academics with requests for submissions, in order to collect author payments. In recent years predatory journals have received a lot of negative media. While much has been said about the harm that such journals cause to academic publishing in general, an overlooked aspect is how much articles in such journals are actually read and in particular cited, that is if they have any significant impact on the research in their fields. Other studies have already demonstrated that only some of the articles in predatory journals contain faulty and directly harmful results, while a lot of the articles present mediocre and poorly reported studies. We studied citation statistics over a five-year period in Google Scholar for 250 random articles published in such journals in 2014, and found an average of 2,6 citations per article and that 60 % of the articles had no citations at all. For comparison a random sample of articles published in the approximately 25,000 peer reviewed journals included in the Scopus index had an average of 18,1 citations in the same period with only 9 % receiving no citations. We conclude that articles published in predatory journals have little scientific impact.

## Introduction

"Predatory journals – authors and readers beware"(Vakil, 2019), is the title of one out of hundreds of similar articles and editorials in the scholarly press, triggered by the concerns of the academic community about the rapid emergence of dubious journals falsely claiming to be scholarly peer reviewed ones. The phenomenon has also been noticed in the general news media (Hern and Duncan, 2018). Much of the attention has been focused on the deceitful behaviour of such publishers, and on a couple of experiments with flawed or nonsensical manuscripts easily passing a non-existent peer review in many such journals. There have, however, been few attempts at empirically measuring the actual impact of the articles published in such journals.

Predatory publishers are companies or individuals who have started new electronic only journals, claiming to fulfil the norms of scholarly peer review and quality control. The primary objective is to collect income in the form of publishing charges (so-called APCs) from authors. Such journals typically spam academics with requests for articles, promising easy and rapid publication, and hunt for scholars willing to join their editorial boards. The exact location of the publishers is often obscure and misleading, with many using names starting in "American" or "Canadian", while actually being published elsewhere. The first warnings about such fraudulent practices were raised in 2008 (Eysenbach, 2008). The phenomenon became more widely noticed due the blacklist of such publishers and journals that librarian Jeffrey Beall started to publish in 2010 (Beall, 2012). The experiment by reporter John Bohannon with a seriously flawed and faked manuscript, which was accepted for publishing by a large number of such journals, further raised awareness of the problem (Bohannon, 2013).

Predatory publishing of journals that pretend to have a peer review system is as a phenomenon in fact part of a broader spectrum of similar web-enabled fraudulent business activities. These include the fake impact factors used by predatory publishers trying to look more credible (Jalalian, 2015), predatory conferences (Cress, 2017) and fake academic degrees (Groelleau et al, 2008).

From a technical and business model viewpoint, predatory journals are also a segment of Open Access (OA) publishing. OA is a disruptive innovation, which is transforming the publishing of academic peer reviewed journals (Björk, 2017). In OA the content is freely available for anyone with web access, and the revenue to the publisher is collected in other ways than via subscription income. The dominating way among commercial OA publishers is via article processing (or publishing) charges, usually abbreviated as APCs. For many non-commercial journals subsidies from scholarly societies, universities or government grants finance the publishing, instead of the authors. During the past two decades OA has slowly but steadily increased its market share of scholarly articles to almost 20 % (Jubb et al, 2017, Piwovar et al, 2018). Articles in Predatory OA journals are not included in these figures which is based on respectable journals indexed in the Scopus journal index.

## Literature review

There have been hundreds of articles about predatory OA but many of these are opinion pieces discussing and lamenting the fraudulent practices of predatory publishers and warning authors from submitting to them. The more systematic or analytical studies can be grouped into a number of thematic categories. The following table names examples of studies falling into these categories, without attempting to be a systematic review of the extant literature.

Table 1. Topical areas for studies of predatory OA journals

| Topical area | Studies |
| --- | --- |
| Characteristic features of predatory OA publishers and journals | Beall, 2012b, Cobey et al, 2018, Bolshete, 2017, Strinzel et al, 2019, Grudniewicz et al, 2019 |
| Role of Blacklists and whitelists | Berger and Ciraselle, 2015, da Silva and Tsiagris, 2018, Olivarez et al, 2018 |
| Number of journals and publishers | Shen and Björk, 2015 |
| Geographical distribution of predatory publishers and authors | Xia et al, 2015, Shen and Björk, 2015, Bagues et al., 2019, Moher et al, 2017 |
| Case studies of predatory publishers and practices | Djuric, 2015, Lukić, 2014, Spears, 2015 |
| Indexing of Predatory journals in WoS, Scopus or DOAJ | Machácek and Srholec, 2016, Demir, 2018, Baker, 2016 |
| Motivation of authors for publishing in predatory journals | Frandsen, 2019, Omobowale et al, 2014, Shehata and Elgllab, 2018 |
| Probes with flawed and nonsensical manuscripts | Bohannon, 2013, Davis, 2009 |
| Studies of article quality | Oermann et al, 2018, McCutcheon et al, 2016 |
| Citation studies | Nwagwu and Ojemeni, 2014, Andersson, 2019 |

The defining characteristics of what constitutes a predatory journal or publisher have over the years been debated. Beall originally used a list of 25 predatory practices and a number of less fraudulent practices as the basis for his blacklist (Beall 2012 b). Bolshete (2017) discusses characteristics of predatory OA journals based on a detailed analysis of thirteen journals. Cobey et al (2018) conducted a systematic review of 38 articles discussing such criteria and found a total of 109 unique characteristics, which they thematically grouped into six categories. Stinzel et al (2019) compared a total of 198 criteria found in blacklists and whitelists and grouped them thematically. At a recent two-day conference devoted to this topic 43 scientists reached the following definition:

""Predatory journals and publishers are entities that prioritize self-interest at the expense of scholarship and are characterized by false or misleading information, deviation from best editorial and publication practices, a lack of transparency, and/or the use of aggressive and indiscriminate solicitation practices." (Grudniewicz et al, 2019)

Beall's blacklist gained a lot of attention worldwide when it was launched. Several authors criticized the criteria and Beall's lack of transparency in using them (Berger and Ciraselle, 2015, da Silva and Tsigaris, 2018), and some publishers included in the blacklist even resorted to lawsuits against him. Eventually in 2017, Beall discontinued the list, but archived versions can still be found on the web. Based partly on his work, the information service Cabell's launched a second blacklist in 2017. Cabell uses a list of 65 criteria for inclusion in its index, and also provides an appeals procedure for publishers who want to contest inclusion. Compared to Beall's list a positive aspect of Cabell's is the listing of all individual journals, not just publishers. The biggest drawback is that the list is not openly available, but requires a subscription (Strielkowski, 2018).

A different kind of solution, less prone to critique, is the use of inclusive whitelists of open access journals which fulfill peer review quality criteria. Such lists could then be used to inform quality assessments much in the same way as Journal Citation Reports. The Scholarly Open Access Publishers Association (OASPA) for instance only accepts as members publishers who meet a number of quality and transparency criteria defined by the association. And the Directory of Open Access journals (DOAJ), in a systematic attempt to prune out predatory OA in 2015-2016, tightened its inclusion criteria and required that all earlier indexed journals applied anew for inclusion (Baker, 2016).

Several authors have reported on the practices of individual predatory publishers and journals (Djuric 2015, Lukić, 2014). A particularly intriguing case was the Canadian Journal *Experimental & Clinical Cardiology*, which was purchased by an unknown company in order to capitalize on the journal already being included in the Web of Science and having an impact factor (Spears, 2014). The journal was converted to open access and started charging an APC of 1,200 US Dollars, and skipped peer review altogether. From 2013 to 2014 the number of published articles in the journal increased from 63 to over 1,000 articles (Shen and Björk, 2015).

An important issue has from the start been to get a perspective of the size of the problem, both concerning the number of predatory publishers, journals and articles published in them. Using Beall's list as basis for data collection has been problematic since he maintained two lists, one of single journal publishers and one of publishers with several journals. The latter list however had no information about the number of journals of these publishers or their titles. In some cases publishers had created portfolios of hundreds of journals. Hence, the total number of predatory journals was for a long time unknown. Shen and Björk (2015) tackled this by first studying all entries in the multiple journal publisher list to determine the number of journals for each. The total number of journals found was 11,873. Based on sampling they estimated that in 2014 the publishers in Beall's list published an estimated 420,000 articles. Cabell's blacklist of predatory journals listed 10,332 journal in November 2018 (Kanto-Karvonen, 2019).

The geographical distribution of authors who have published in predatory journals has for instance been studied by Xia et al (2015). They found that they are mostly young and inexperienced academics from developing countries. Dominating countries of origin were India, Nigeria and Pakistan. Shen and Björk (2015) found that two thirds of authors originate from Asia and Africa. There are however recent reports that also scholars from leading developed countries to some extent publish in predatory journals. Bagues et al (2019) found that 5 % of Italian academics seeking promotion had published at least one article in a journal included in Beall's list. Moher et al, in a study of 1,907 biomedical articles in predatory journals, found that more than half of the articles stemmed from authors from high and upper middle income countries according to the classification of the World Bank.

Due to the difficulty of identifying predatory journals among OA journals that charge APCs some predatory journals have been listed in indexes such as Scopus and even Web of Science, and more so in DOAJ prior to 2016. Machácek and Srholec (2016) found that of the journals included in Beall's list, 3,218 were indexed in Ulrichsweb and 405 in Scopus. Demir (2018) found that of the 2,708 journals in Beall's list that they studied, only three were in also the Web of Science and 53 in Scopus. Reasons for this could be that either a journal which clearly is predatory has by mistake been included in the more rigorous indexes, or that a journal has mistakenly been included in Beall's list, when it should not.

Frandsen (2019) reviews some earlier studies of the motivation of authors for publishing in predatory journals. She points out that there are two different categories of academics who have published in such journals. Those who are uninformed about the nature of the journals, and those who are aware of the situation but choose to publish in them as a low-barrier way to get published for expected career gains. She also notes that it is very difficult to empirically study their motivations for instance using surveys, since those belonging to the latter group would be unlikely to admit to their motives. Omobowale et al (2014), in interviews with 30 academics from Nigeria, found that a major reason for publishing in such journals was to satisfy the "international publishing rule" at all costs. Shehata and Elgllab (2018) surveyed and interviewed Egyptian and Saudi Arabian scholars who had published in 18 predatory journals. They found that easiness and speed were major factors influencing the submission.

There have been two probes with flawed manuscripts that have gained a lot of publicity also in the popular press. Davis (2009) reported that he and a colleague had submitted a grammatically correct but nonsensical manuscript generated by a software program to a predatory journal, which was rapidly accepted for publishing, under condition of payment of a publication charge of 800 USD. A manuscript containing major methodological errors and other weaknesses sent out by journalist John Bohannon (2013) was accepted by 157 target journals and rejected only by 98. While such experiments demonstrate that the peer review practices are often so deficient or totally lacking that just about any sort of paper could be accepted for publishing in many of these journals, they tell little about the scientific quality of the average papers in these journals.

There have been only a couple of studies which have tried to investigate the scientific quality of the articles published in predatory journals via a post-publication peer review of the full-text articles. Oermann et al (2018) studied a random sample of 358 articles published in predatory journals in Nursing. They found that 50 % of the articles presented content that was useful for nurses and that 32 % had flaws such as lack of human subjects review or in-correct research design. In the overall assessment of the research group 171 articles were rated as poor, 169 as average and 13 as excellent. A particularly interesting finding is however, that the authors found that only 5 % of the articles in the sample were judged to be potentially harmful to patients or others, although many of the articles represent poor scholarship. McCutcheon et al (2016) post peer reviewed 25 psychology articles in predatory journals comparing them to 25 articles in regular journals and found significant differences according to five criteria.

We found two previous studies looking at citations to articles in predatory journals. Nwagwu and Ojemeni (2014) did a bibliometric study of 36 biomedical journals published by two Nigerian predatory publishers. For 5,601 articles published in 2007-2012, they found a total of 2,772 citations in Google Scholar in 2014. Andersson (2019) reports in a blogpost on citations in WoS, Science Direct or *PLOSONE* to articles published in seven predatory journals. He notes in particular the high incidence of WoS citations to two of the journals (25 and 37 % of their articles had at least one citation). The identity of the journals is however not revealed and the method description is quite vague in his blog post.

The probable cause of the lack of studies of citations is that since predatory OA journals are usually excluded from the major citation indexes (ie. WoS and Scopus) there is no easy data to use on the journal level. Nevertheless, a citation study would answer the important question of whether articles in predatory journals have any measurable impact on the work of other scientists (which could be harmful only in the case of articles presenting false results, not in the fact of mediocre but methodologically sound research).

**The research question of this study was hence to study how frequently articles published in predatory OA journals are cited, as a proxy for the influence of these articles on the research of others.**

# Methods

In order to study the research question in a meaningful way a random sample of articles published in predatory journals is needed. We chose to study articles published in 2014, since enough time had elapsed in 2019 for them to potentially accumulate citations. With the very low numbers of citations that we expected two years would for instance have been a too short window. In five years a useful article should accumulate also second generation citations, where the citing scientist had identified the article via a citation in another publication, rather than finding the article via some direct search using key words.

In bibliometric citation studies the standard practice has been to use Web of Science data and in recent years also data from Scopus. This is because both of these indexes in addition to articles also keep track of citations, which makes studies with large data sets possible. For the purpose of this study, the use of citation data from either of these indexes was however not feasible, since only a small fraction of "predatory" journals are included in either index (see for instance Machácek and Srholec, 2016). Instead the only realistic way was to use Google Scholar (GS). In GS once a scientific publication has been found, the number of citations to it in other GS indexed publications is also shown.

As a basis for choosing articles we used Cabell's blacklist of journals "as potentially not following scientific publication standards or expectations on quality, peer reviewing, or proper metrics". This is a commercially maintained list which has succeeded the earlier Beall's list. Cabell's list has a major advantage compared to Beall's earlier list, in that it actually directly names over 10,000 journals. Cabell's service is subscription based, but the publisher kindly agreed to grant us free access for our research purposes. In fact after we found selecting random articles from the normal web based search interface cumbersome, they also provided an Excel list of all the journals, which much facilitated the selection process. Our aim was to collect a usable set of 250 journals which had published articles in 2014. The number 250 was originally set when we started identifying journals from the web search interface, which shows 40 journals at time, and we chose not to change the sample size later when we used randomized journals from the excel data. For each of the identified journals the journal's web site was searched for an article published in 2014. Next, the sample articles were checked using Google Scholar. The number of citations, the lack of citations, or not being found at all on Scholar were noted down.

A minor drawback of this method is that the sample is not directly random over all articles published in Cabell listed journals published in 2014. This is because the number of articles published varies from journal to journal. There should ideally be a higher probability of an article from a bigger volume journal of being included in the sample. Achieving this would, however, have been extremely tedious, since that would have meant first hand-counting the number of articles published in 2014 by over 10,000 journals from their websites, in order to take account of this fact in producing the sample.

We also wanted to study the number of Google Scholar citations to articles in a control set of non-predatory journals from the same period. Our primary control group consists of the set of articles published annually in Scopus indexed peer reviewed journals. Scopus imposes quality checks on journals included in the index but is not as restrictive as Web of Science in admitting new journals. Currently there are around 25,000 journals listed in Scopus publishing approximately 2,5 million articles annually. We were provided with a randomly picked set of 1,000 articles published in 2014 by Elsevier's ICSR Lab which facilitated our data collection.

From this set we took the first 250 articles and performed manually the same Google Scholar searches as for the predatory journal articles. Additionally the data provided by the ICSR lab also included the number of citations until now in Scopus journals for each article, which is also included in the results reporting. We expected these citation numbers to be lower than the Google Scholar ones.

Another useful comparison is to citations for such OA journals, which follow standard peer review practices. OA journals are a good comparison group for our study, since any positive effects on citations from the articles being openly available on the web is the same as for the predatory articles. Several studies have claimed such a positive OA citation effect (Lewis, 2018). Most of such journals are indexed in the Directory of Open Access Journals (DOAJ), although there are also many such journals not included, in particular from different regions of the world. For practical reasons we chose to focus on the subset of DOAJ journals which also are indexed in Scopus. By comparison of title names with the current DOAJ list of journals we were able to extract 107 articles published in full OA journals from our set of 1,000 Scopus articles. This set includes 14 articles from the megajournal PLOS ONE, which in 2014 published around 30,000 articles. This is pretty much in line with the overall share of PLOS ONE articles of all OA journal articles. For this set we also got the Scopus citations with the data.

Full OA journals are either older subscription journals, which when they started to make available an electronic version made that OA. Or they are journals which have been founded as OA journals from the start. We feel that "Born OA" journals offer a particularly useful comparison, since they are usually younger and have had to build up their scholarly reputation from scratch. Such journals include for instance many journals from specialized OA publishers like PLOS, BMC, Hindawi and MPDI.

For this purpose a second OA control group could be established by using data from a separate on-going study in the research group at the Hanken School of Economics. In that study, OA-journals have been identified which are indexed in either the DOAJ or the ROAD indexes of OA journals, and at the same time also in Scopus. Furthermore these journals have in an on-going research project been manually inspected and classified into either older subscription journals which have made the electronic version Open Access (converted journals), or journals which from the start have been launched as electronic OA journals (Born OA journals). We obtained the data for 250 born OA journals and then chose one article from each of these published in 2014.

It is important to note that two of the samples are direct random samples from the population of all articles studied, while for two of the samples a random sample of journals has first been produced and after that one article has been extracted for each journals. This means that journals with big publishing volumes are proportionally represented in the samples based on direct data from Scopus, while they only have a small chance of being included in the predatory or born-OA set.

## Results

An important secondary finding of this research was that in order to locate a sample of 250 predatory journals from Cabell's list, which had published articles in 2014 that are still findable on the web, required investigating 595 journals. It turned out that the quality of the sites of the publishers varied a lot, from simple info-pages to highly functioning search functions and archives with visually thought-out appearances. The sample chosen from Cabell's list contained many journals, whose web addresses in the index didn't work. Also in some cases the data security software on the researcher's computer prevented access. Some of the journals and their articles were findable, but had no articles published in 2014. The reasons for excluding journals is shown in figure 1.

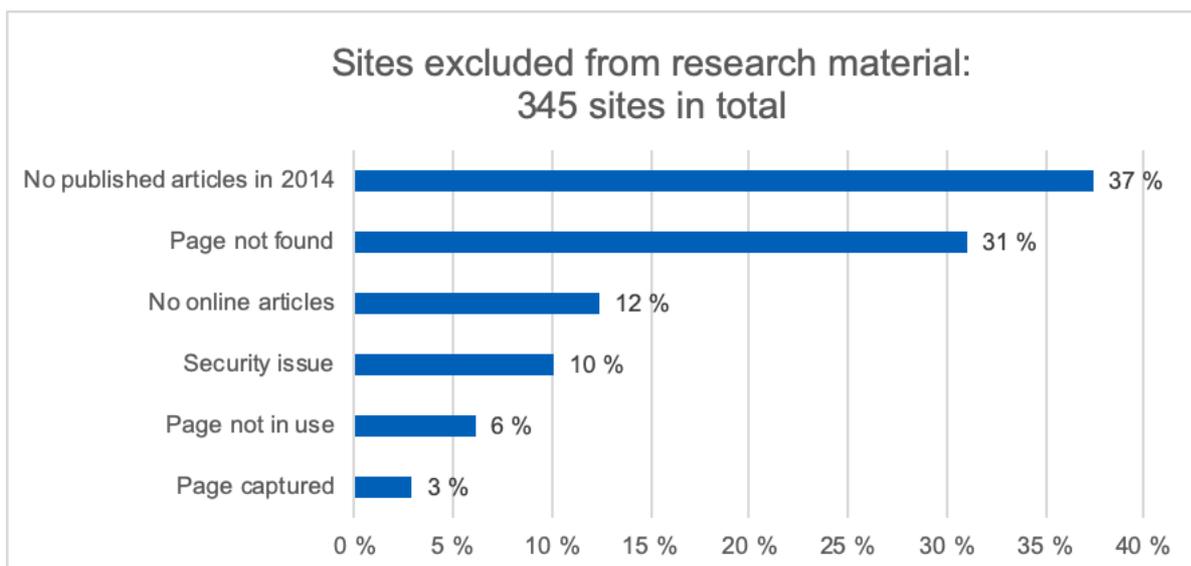

Figure 1 Reasons for excluding journals in the initial random set.

The distribution of citations to the predatory articles was highly skewed, we noticed that almost half of the citations where to only ten articles with between 13-43 citations. Since several critics have noted that Beall's list (and by extrapolation potentially also Cabell's list) may contain journals that have falsely classified as predatory, we further doublechecked the ten journals in questions. In our judgement at least four among them should not have been included in the results. For instance the highest cited article in our sample (41 citations) was published in a WoS indexed journal with an impact factor of 5,5. The journal's articles are also included in the leading medical

PMC repository. Excluding the four articles we had identified dropped the number of citations by 124.

The distribution of the 513 citations to the remaining articles is shown in figure 2. There were no citations to 60 % of the articles, and the mean number of citations per article was 2,1. The median was zero.

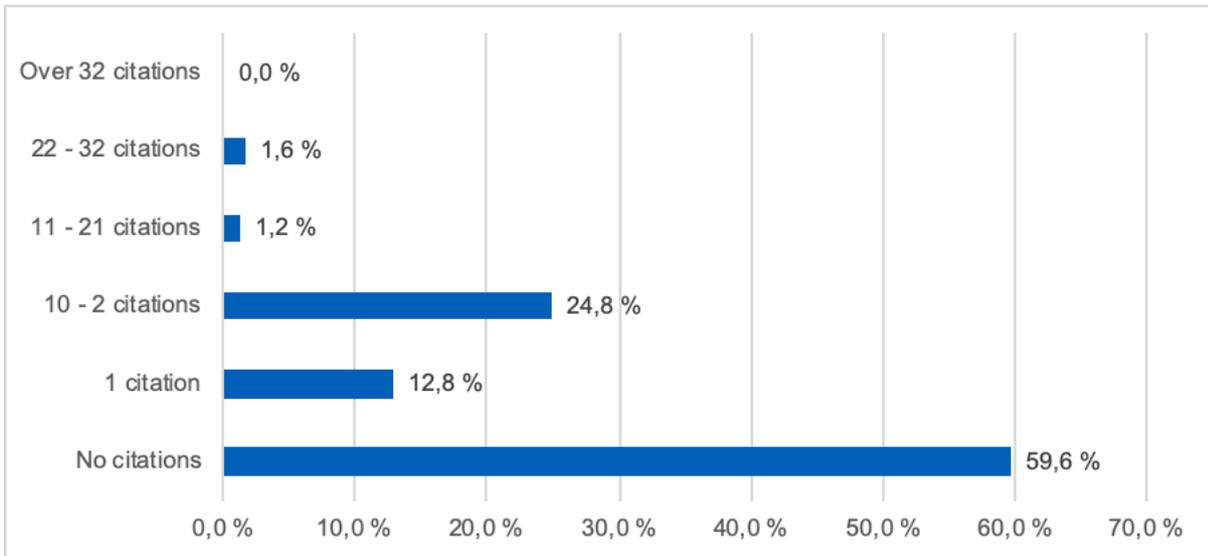

Figure 2. Citations since 2014 to the articles in predatory journals

We also studied in detail where the ten articles which had received citations in the range of 10-32 were cited. The 212 citations had been in 40 journals, of which 7 were on Cabells' list. Finally, we analysed the reference lists of the citing articles, which contained a total of 809 references. Of those, 82.5% were not on Cabells' list.

The main results for all for all sets of articles studied are shown in table 2 below.

Table 2. Descriptive statistics for the citations to articles published in 2014 in Predatory OA journals as well as to three control groups. The 95 % confidence intervals are indicated in parentheses after the averages.

|  | Google Scholar citations | | Scopus citations |
|---|---|---|---|
|  | Share with zero citations | Average number of citations | Average number of citations |
| All Scopus indexed articles | 9 % | 18,1 (± 2,7) | 12,6 (± 2,1) |
| Scopus indexed articles published in Open Access journals | 7 % | 16,5 (± 3,2) | 10,6 (± 2,55) |
| Articles in Born OA journals indexed in Scopus | 18 % | 12,4 (± 3,2) | |
| Articles in Predatory journals | 60 % | 2,6 (± 0,7) | |

The difference to articles in Scopus (which indexes the vast majority of the scholarly peer reviewed articles published in English) is very clear. Scopus articles receive on average around seven times as many citations in Google Scholar as Predatory journal articles. Also the number of articles with zero citations in GS is only 9 % for Scopus articles compared to 60 % for predatory articles. The two control groups of "quality assured" OA journals included in Scopus also have much higher citation statistics. The difference between these two groups can largely be explained by the sampling methodologies used to construct the comparison sets. It is also noteworthy that articles in OA journals in general are almost as frequently cited as scholarly journals in general.

An interesting side result of this study is also that it shows that the number of citations in Google Scholar for this time window is around one and half time the number of citations in Scopus, for articles which are indexed in both.

Kousha and Thelwall (2017) have published an interesting article were they discuss the impact of scientific books and articles in non-scholarly works such as Wikipedia, teaching materials and clinical medical guidelines. These are channels were references to and use of material from faulty articles on contagious subjects published in predatory journals could potentially be very harmful. The authors found for instance that there were Wikipedia citations to only 5 % of over 300,000 Scopus indexed articles published between 2005 and 2012. Teplitskiy at al (2017) studied in particular whether open access to the articles increased the probability of an article in high quality journals in Scopus being referenced in Wikipedia and found that the odds increased by 47 % compared to articles in paywall journals. We performed a small scale test with the 17 most highly GS-referenced articles in our predatory article sample and found no Wikipedia references.

## Discussion

The effects of predatory publishing can be analyzed in different ways. While a couple of experiments with flagrantly flawed or nonsense manuscripts have demonstrated that they can pass the non-existent peer review in many predatory journals, they prove little about the average quality of articles published in such journals. For instance the study by Oermann et al (2018) that did a post-publication review of articles showed that in only a few cases articles in the sample were judged to be potentially harmful to patients or others, although many of the articles represent poor scholarship. In the past, many such articles would probably have been published in local print based scholarly journals, or as grey literature such as departmental working papers etc. Also in some cases the articles could eventually have passed the more rigorous peer review of conventional scholarly journals, but the author opted for the fast-track and easier option of a predatory journal.

Another important issue is the effects of predatory publishing on different stakeholders as for instance discussed by Eve and Priego (2017) or Frandsen(2019). Firstly there is the effect of the articles on the science of other researchers. Our measurement of citations seeks in particular to address this. Secondly there is the possibility of faulty results or claims being picked up by the general public and spread via social media. Thirdly there is the effect on the integrity on the academic evaluation system, such as filling of academic positions, or the allocation of grants. And lastly there is the negative side effects of the bad publicity of predatory journals on the development of credible open access journals, affecting their ability to attract good submissions.

We feel that too much focus in the discussion of predatory publishing has been put on the negative effects on academic evaluation practice. Several of the countries from which a majority of authors in predatory journals stem also rank very badly in terms of international corruption comparisons. Thus giving authors undue credit for articles in predatory journals is likely to be only one kind of corruption in play in such countries.

What our results demonstrate clearly is that the average predatory journal article has very little effect on the research of others and that it probably also has very limited readership among academics.

## Conclusions

The articles published in so-called predatory OA journals can be grouped into two main categories. Part of the articles present results which are scientifically valid, but which would probably not have passed the review of more selective international journals. The reasons are that the writers English or skills of writing up articles are not sufficient, that the results are more or less replication of earlier research, the review of earlier research is not thorough enough or the topics may be of only local interest etc. The authors may also have opted for a predatory journal because of earlier rejections or because of the need for rapid publication or publication in a journal with an international label. If such research is read and possibly even cited by other researchers there is little harm done, and in some cases it can even be useful.

More problematic are studies, which have clear methodological flaws and draw the wrong conclusions. The manuscripts submitted in the probes that have received a lot of publicity are extreme cases. But also in such cases there is little harm done if nobody reads and in particular makes use of such results. As Donovan (2017) has pointed out in a comment to the report by Moher et al (2017): "Predatory journals: Research that isn't read doesn't exist". The biggest risks with articles in predatory journals would be articles on highly contagious topics (ie climate change, harmful side-effects of vaccination), which might be picked up by advocacy groups and spread via social media to promote the interests of such groups.

Overall we found few citations to Predatory journal articles, even though we used a five-year window and Google Scholar, which typically finds more citations than Scopus and in particular Web of Science. More than half of the articles in our sample of predatory journals had no citations at all. The few articles we found with more than 10 citations turned in some cases out to be in journals with credible peer review, which seem to have been mistakenly, classified as predatory journals.

Our study suggests a number of possible directions for further research. One obvious one is to use more control groups. In addition to comparing to highly quality journals indexed in Scopus/WoS, where journals from major publishers based in the Anglo-American Countries dominate, such comparisons could also include more journals from the same parts of the world where the authors in predatory journals predominantly come from, for instance using DOAJ journals not indexed in Scopus. An interesting question is also how frequently articles in predatory journals are cited in social media such as Twitter and Facebook. Zheng et al (2019) have discussed the frequency of such citations for scholarly journal articles in general. Another route is to continue the studies reviewing the actual quality of the articles published in such journals using expert evaluations.

Perhaps the single most negative aspect of predatory publishing is that it has cast a shadow on the development of more responsible Open Access Publishing and has possible slowed down its development. Many academic authors have unnecessarily equated open access and APCs with the lack of peer review and quality.

## Acknowledgements

The initial data collection for this study was carried out by Sari Kanto-Karvonen as part of her master's thesis in information science at the Tampere University. Kanto-Karvonen later expanded the work to include more data sets, as per the requirements of this study. J. Tuomas Harviainen was the thesis supervisor. Bo-Christer Björk proposed the research topic and acquired the research material. All three participated in the planning of the study and in the writing of this article. We would like to thank Cabell's for providing us with the data which was essential for selecting the sample of predatory journals used in the study. This work also uses Scopus data provided by Elsevier through the ICSR Lab. The data about the born-OA journals used as one of the control groups was obtained from Mikael Laakso.